\documentclass[universe,article,submit,moreauthors,pdftex]{Definitions/mdpi} 
\firstpage{1} 
\pubvolume{1}
\issuenum{1}
\articlenumber{0}
\pubyear{2021}
\copyrightyear{2020}
\datereceived{} 
\dateaccepted{} 
\datepublished{} 
\hreflink{https://doi.org/} 

\usepackage{natbib}
\usepackage{graphicx}	
\usepackage{amsmath}
\usepackage{amssymb}
\usepackage{hyperref}

\Title{Impact of AGB stars on the chemical evolution of
neutron-capture elements}

\TitleCitation{AGB and CE of neutron capture elements}

\Author{Gabriele Cescutti $^{1,2,3,4}$\orcidA{}, Francesca Matteucci $^{1,2,3,4}$\orcidB{}{}}

\AuthorNames{Gabriele Cescutti, Francesca Matteucci}

\AuthorCitation{Cescutti, G.; Matteucci, F.}

\address{
$^{1}$ \quad Dipartimento di Fisica, Sezione di Astronomia, Università di Trieste, Via G. B. Tiepolo 11, 34143 Trieste, Italy\\
$^{2}$\quad INAF, Osservatorio Astronomico di Trieste, Via Tiepolo 11,  I-34143 Trieste, Italy\\
$^{3}$\quad INFN, Sezione di Trieste, Via A. Valerio 2, I-34127 Trieste, Italy\\
$^{4}$ \quad IFPU, Istitute for the Fundamental Physics of the Universe, Via Beirut, 2, I-34151 Grignano, Trieste, Italy\\
}

\corres{Correspondence: gabriele.cescutti@inaf.it}

\abstract
{In this paper we discuss the impact of the s-process nucleosynthesis in Asymptotic Giant Branch stars on the enrichment of heavy elements. We review the main steps made on this subject in the last 40 years and discuss the importance of modelling the evolution of the abundances of such elements in our Milky Way. From the comparison between model results and observations, we can impose strong constraints on stellar nucleosynthesis as well as on the evolution of the Milky Way.}
\begin{document}
\section{Introduction}
Over 70\% of the chemical species in Nature are constituted by neutron capture processes and more than 50\% of the neutron capture elements that we find nowadays in the Sun were formed in the interior of asymptotic giant branch (AGB) stars. 
It is not trivial to calculate precisely the production of such elements, and it is necessary to know  the detailed nucleosynthesis in AGBs extended u-process to the heaviest stable nuclei.{\bf  The majority of elements with $60 \le A \le 90$ are produced in massive stars and represent the "weak" s-process component, while for $A > 90$  the elements are formed in AGB stars and they represent the "main"  s-process component. Finally, very heavy elements such as $^{208}Pb$ represent the "strong" s-process component, and they are formed AGBs, but of low metallicity. The s-process elements are created by neutron capture on Fe seeds on timescales longer than the $\beta$-decay in the nucleus, and the neutrons are released by two reactions, $^{13}C(\alpha,n)^{16}O$ and $^{22}Ne(\alpha,n)^{25}Mg$. This is a sort of secondary process because it depends upon the abundance of the seeds. However, it would be better to say that is a complex  process since the neutron flux in turn depends on the stellar metallicity and because of the presence of poisons (light elements that capture the neutrons which will no more contribute to the s-process), such as $^{14}N$.  For example,  the formation of Pb is favoured by a low stellar metallicity. In fact, when the abundance of Fe seeds is low, it is easier to form  very heavy s-process  elements, because the neutron capture is concentrated on a lower amount of seeds. For the same reason, the light s-process elements (Sr, Y, Zr) are favoured by high metallicity.} This kind of calculations was pioneered by \citet{Gallino88,Gallino98} and was also deeply studied by \citet{Busso95,Busso01}. Although it is possible to foresee the behaviour of the abundances of chemical elements just  by considerations based only on stellar nucleosynthesis, we need to adopt self-consistent chemical evolution models if we want to take into account the complexity of the processes and obtain detailed results to be compared to the observations.  In such models we can take into account also the other production channels of neutron capture elements such as the production by the r-process,
\end{paracol}
weak r-process (possibly in the future also the contribution of less studied rare processes such as i-process, p-process or $\nu$p-process), as well as the iron abundance evolution, which is the tracer of metallicity in stars.  
Due to the complexity of the various sources and their strong metallicity dependency, the chemical evolution of neutron capture elements is more complex and demanding than the one of $\alpha-$elements and iron peak elements. On the other hand, thanks to the careful studies of these elements, we can impose strong constraints on stellar nucleosynthesis and  formation and evolution of our Galaxy. 
In the present paper, we will present a selection of results obtained by adopting chemical evolution models including detailed nucleosynthesis of s-process elements originating from AGB stars, during the last 40 years.

 \section{Galactic Chemical Evolution}
Models of galactic chemical evolution are aimed at predicting how the abundances of chemical elements do evolve in the gas in galaxies. The main ingredients necessary to build such models are the
stellar birthrate function, in other words the star formation rate (SFR), which is normally assumed to be a continuous function in the Milky Way such as the Kennicutt (1998) law, and the initial mass function (IMF). Then a fundamental ingredient is stellar nucleosynthesis, namely the amounts of chemical elements formed inside stars and restored into the interstellar medium (ISM) by means of stellar winds and supernova (SN) explosions. Supernovae can arise from massive stars ($M> 8M_{\odot}$) which do explode by core-collapse giving rise to Type II, Ib and Ic supernovae,  and have short lifetimes. These stars produce the bulk of $\alpha$-elements (e.g. C, O, Ne, Mg, Si, S, Ca) and part of Fe. On the other hand, supernovae Type Ia arise from exploding white dwarfs in binary systems and they have longer lives and do produce the bulk of Fe. However, also low and intermediate mass stars ($0.8< M/M_{\odot}\le 8$) during the AGB phase can be important element producers, in particular they are responsible for $^{14}N$, $^{12}$C and for the bulk of the s-process component in neutron capture elements, as mentioned in the Introduction. {\bf Concerning r-process elements they should arise from massive stars ($M> 8M_{\odot}$), either from core-collapse SNe or merging neutron stars, and the roles of these sources are still under debate.}
We can divide the galactic chemical evolution (GCE) models into two categories: analytical and numerical models, as we will describe in the following.

\subsection{Analytical chemical models}

The simplest and oldest analytical model is the {\it Simple Model} (\citet{Tinsley80}) which assumes that the system is a closed box (no infall nor outflow), the IMF is constant in time, the initial gas has a primordial chemical composition and there is instantaneous mixing of the stellar products with the ISM. In addition, there is the hypothesis of instantaneous recycling approximation (IRA), which states that all stars below 1$M{\odot}$ never die, while those above  1$M_{\odot}$ die immediately. These assumptions lead to simple solutions for both primary (originating directly from H and He) and secondary  (originating from metals) elements.
The solution of the Simple Model for a primary element is:
\begin{equation}
    Z=y_Z  ln (\frac{1}{\mu})
\end{equation}
where Z is the global metal content and $y_Z$ is the so-called yield per stellar generation and $\mu=M_{gas}/M_{tot}$ is the fractional mass of gas in the system.
The solution for a secondary element, such as $^{14}$N and $^{13}$C is:
 \begin{equation}
    X_S =1/2 {y_S \over y_Z Z_{\odot}} Z^2
\end{equation}
where  $X_S$ is the mass fraction and $y_S$ the yield per stellar generation of a secondary element $S$, which originates from the primary element $Z$.
Clearly, the adoption of IRA in the  Simple Model is incorrect if one wants to model the evolution of elements  mainly produced in long living stars, such as  s-process elements. 
In such cases, numerical models are necessary to take into account the stellar lifetimes, and the results they produce are quite different from those of the Simple Model, as we will see in the next Sections.
    
\subsection{Numerical chemical models}
Numerical models take into account in detail the stellar lifetimes and therefore are particularly important to model elements produced by AGB stars and Type Ia supernovae. These models relax IRA, as well as the closed-box assumption, and allow for infall or outflow of gas in the studied galaxy. The numerical models follow the evolution of single elements and take into account the nucleosynthesis  products of stars of all masses in great detail. 
In the following,we will describe how the evolution of s-process elements has been interpreted and modeled by different authors, with particular attention to the contribution from  AGB stars.


\section{Truran (1981)}

\begin{figure}[H]	
\widefigure
\includegraphics[width=14 cm]{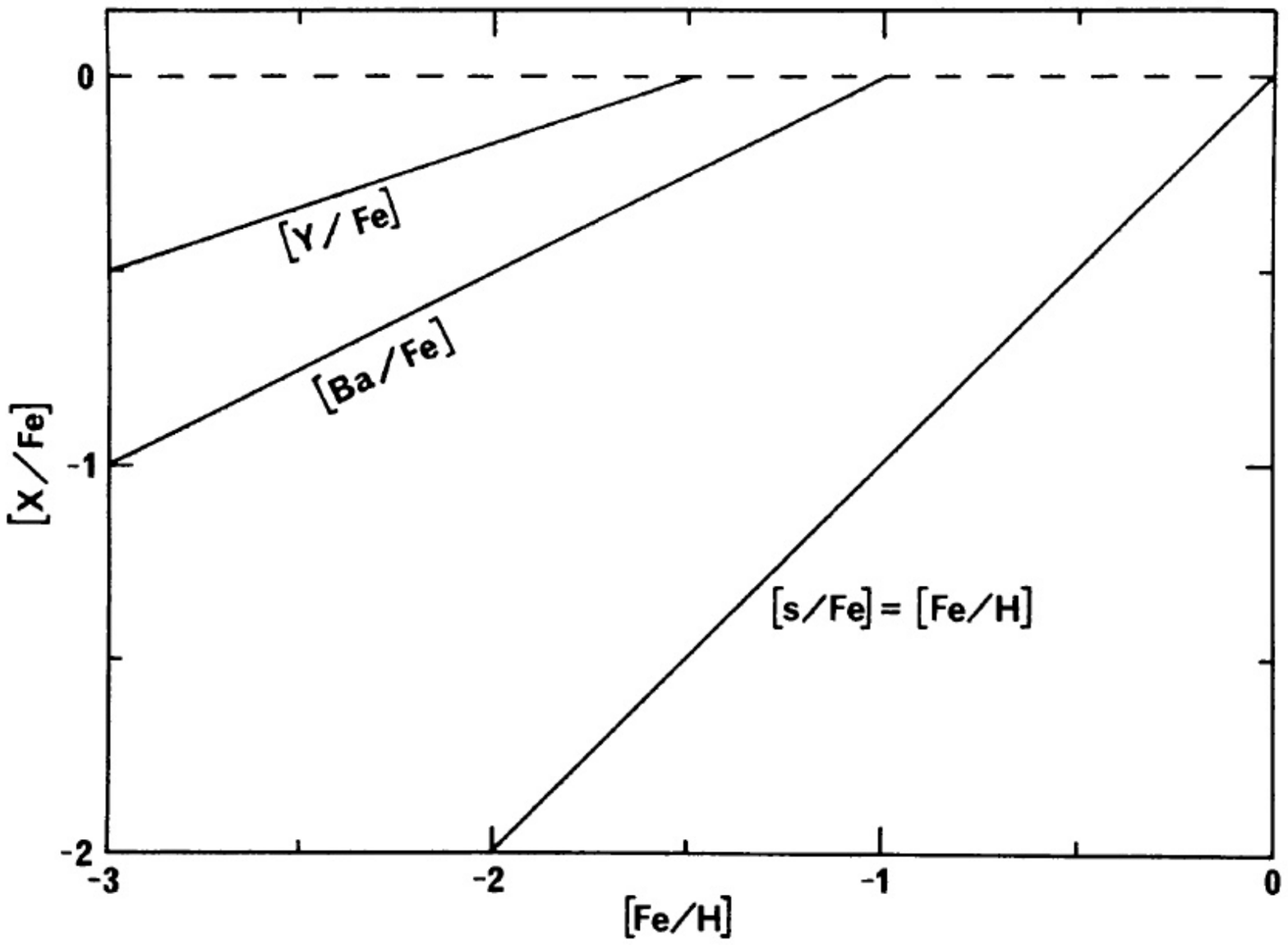}
\caption{Observed behaviour of the ratios of Y and Ba relative to Fe together with the expected trend for purely secondary elements ([s/Fe]) relative to Fe. Image reproduced with permission
 from \citet{Truran81}, copyright by the authors.}
\label{fig1}
\end{figure}

We start with the pioneering results shown in \citet{Truran81}. Although he did not consider a real chemical evolution model nor a detailed nucleosynthesis of the neutron capture elements, nevertheless, he used fundamental concepts of chemical evolution applied to the knowledge of the s-process nucleosynthesis. 
In particular, the fact that 
the s-process nucleosynthesis is expected to behave as a secondary process. So, as described in the previous Section, being the secondary production dependent on the abundance of metal seeds (in particular iron), is expected to increase quadratically with the iron abundance,  and therefore the ratio of the secondary element relative to the seed should be linear, as shown in Figure \ref{fig1}. However, it should be noted that such a behaviour for secondary elements is only  valid in the framework of the Simple Model of chemical evolution, whose solution, both for primary and secondary elements, is obtained  with IRA. {\bf Therefore, it is expected that in a realistic situation, where the complexities of nucleosynthesis and stellar lifetimes are taken into account, the elemental abundances should not follow the predictions of the Simple Model. }

 Clearly, this simple assumption  was not compatible with the behaviour suggested by the first observations of neutron capture elements in metal-poor stars, which showed a different behaviour of  these elements relative to Fe, as shown  also in Figure \ref{fig1}.  Although \citet{Truran81}  made his considerations in the framework of the Simple Model, he envisaged the  right solution to explain the early production of neutron capture elements which should have originated from a different process, the r-process. This was indeed correct and started a new line of thought. Clearly, the evolution of s-process elements relative to Fe as a function of Fe abundance along the entire Galactic lifetime cannot be explained by the Simple Model only but it needs the consideration of stellar lifetimes, as we will see in the next paragraphs. In particular, the stellar lifetimes are particularly important for s-process elements which mainly originate from long living low mass stars (1-3$M_{\odot}$).

In any case, Truran's interpretation suggested the scenario we follow now about the production of neutron capture elements in the early stages of evolution of  our Galaxy.

\section{Travaglio et al. (1999)}

In \citet{Travaglio99}, they studied for the first time the chemical evolution of the Galaxy tracing the  elements  Ba and Eu, using a detailed evolutionary numerical model, relaxing the IRA, and suitable for reproducing a large set of Galactic (local and non local) and extragalactic constraints. Input stellar yields for neutron-rich nuclei were separated into their s-process and r-process components. 
In that work, it was considered a very detailed production of s-process elements in thermally pulsing AGB stars of low mass; in particular, the combined action of two neutron sources was taken into account: the dominant reaction $^{13}$C($\alpha$,n)$^{16}$O, which releases neutrons in radiative conditions during the interpulse phase, and the reaction $^{22}$Ne($\alpha$,n)$^{25}$Mg, marginally activated during thermal instabilities \citep{Gallino98}.{\bf It is worth noting here the fundamental role played by the third dredge-up in the s-process nucleosynthesis, because it drives the formation of a $^{13}C$ pocket, and it is a primary process since $^{13}C$ is created from the original H and He.} {\bf The r-process yields adopted in this work, were not a prediction of stellar nucleosynthesis models and were derived as the difference between the solar abundance and its s-process contribution, given by their GCE model at the time of formation of the Solar System. This approach has became a quite standard procedure. However, it implies a single primary r-process component. It has been shown by comparison with metal poor stars that this approach is valid only for elements heavier than A=130 \citep[e.g. Ba and beyond][]{Sneden03,Sneden08}; already \citet{Travaglio04} invoked the need for a new process that together with s-process from AGB and a primary r-process is 
able to provide an additional contribution to light neutron capture elements (Sr, Y, Zr). The astrophysical conditions that would create this additional contribution are still unknown, and both an s-process-like or an r-process-like mechanism was
found to reproduce the abundance pattern between Sr and Ag
observed in many of the most extreme metal poor stars \citep{Montes07}. 
The s-process production in massive stars (weak s-process) was shown to be inefficient at low metallicity \citep{Raiteri92}, so most authors have focused on the existence of a possible second r-process \citep[also called the weak r-process; see][]{Arcones11}. 

\begin{figure}[H]
\widefigure
\includegraphics[width=12 cm]{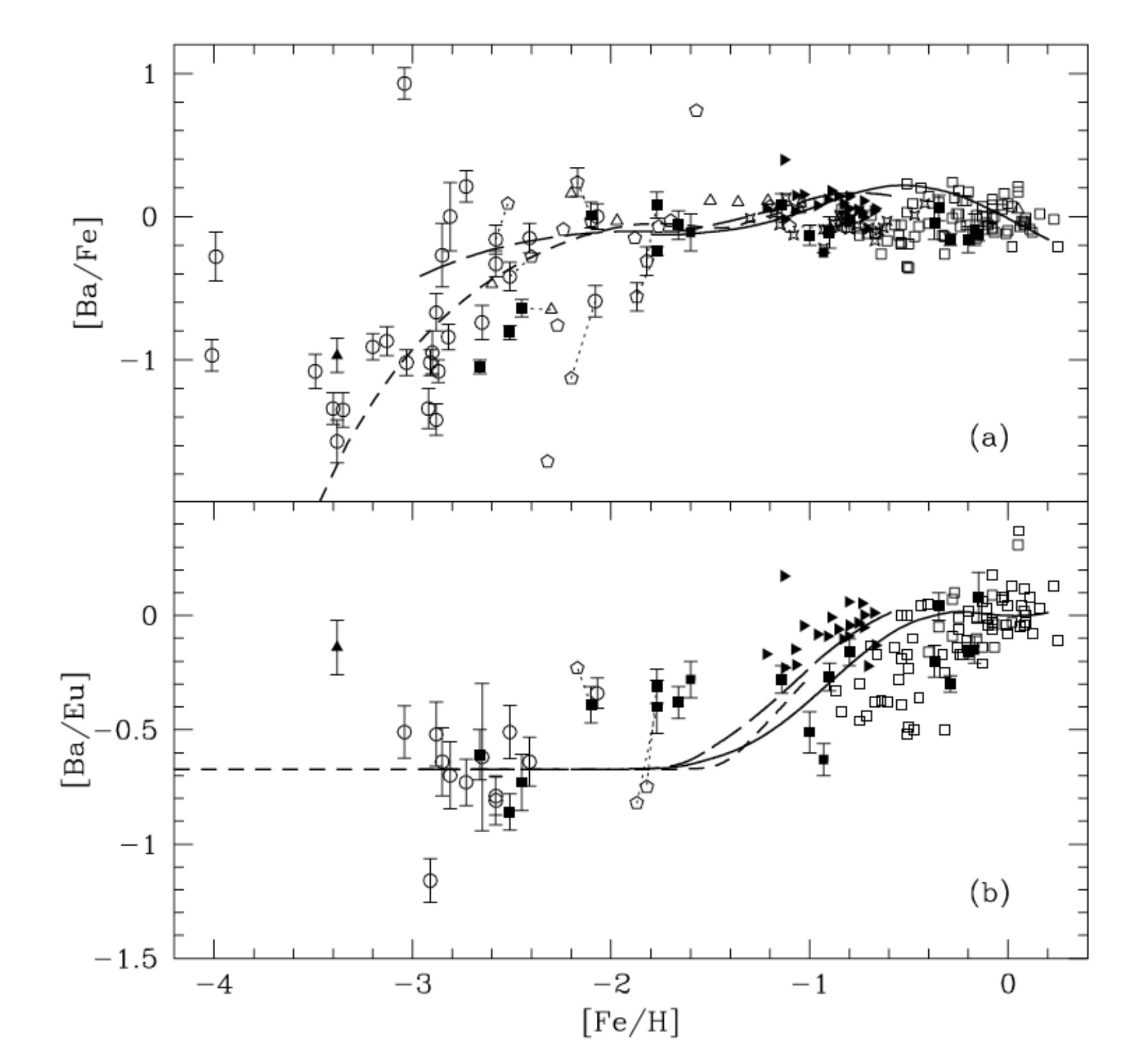}
\caption{[Ba/Fe] vs [Fe/H] in the top panel and [Ba/Eu] vs [Fe/H] in the bottom panel.
 The short dashed line describes the halo, the long dashed shows the thick disk and solid line is for the thin disk. Observational data are from \citet{Gratton94} (filled squares); \citet{Woolf95} (open squares); \citet{Francois96} (pentagons); \citet{McWilliam95} and \citet{McWilliam98} (circles); \citet{Norris97} (filled triangles); \citet{Jehin99} (filled tilted triangles; \citet{Mashonkina99} (open triangles). Thin dotted lines connect stars with different abundance determinations. Image reproduced with permission
 from \citet{Travaglio99}, copyright by the authors.}\label{fig2}
\end{figure}  
However, the recent nucleosynthesis computations by
\citet{Frisch15} and \citet{LimongiChieffi18} showing that rotating massive stars  can support the s-process, have brought a new twist to the interpretation of the neutron capture element abundances of stars at very low
metallicity (see also Sections 7 and 9).}
Galactic evolution results, including both the s- and r-process contributions are compared to the data in Figure 2. As one can see from that figure, the [Ba/Fe] ratio does not follow the behaviour predicted by the Simple model but after an initial linear increase this ratio flattens at higher metallicities. This is the effect of taking into account in detail the stellar lifetimes of the stars producing both Ba and Fe. The Fe is produced mainly by Type Ia SNe which originate from white dwarfs in binary systems, {\bf while Ba is formed mainly in low mass stars}.  The ratio [Ba/Eu] has a different behaviour than [Ba/Fe] since, at variance with Fe, Eu is produced,  as an r-process element, on very short timescales by massive stars. The initial flat behaviour is due to the r-process component of Ba, already suggested by \citet{Truran81}, which is produced on the same timescales as Eu, then the increase of the ratio with increasing metallicity is due to the s-process component of Ba, which is the dominant one and is produced on long timescales by low mass stars.

The resulting s-process distribution was also strongly dependent on the stellar metallicity. For the standard model discussed in that paper, there is a sharp production of the Ba-peak elements around Z$\sim$Z$_{solar}$/4. Concerning the r-process yields, they made a simplified  assumption in which the production of r-nuclei is a primary process and it occurs in stars near the lowest mass limit for Type II supernova progenitors ($ > 8M_{\odot}$). 
With this model, it was possible to calculate the  r-contribution for each nucleus at
different evolutionary times, and in particular at the formation of the Solar System. 
Their results were also compared to spectroscopic abundances of elements from Ba to Eu at various metallicities (mainly from F and G stars), showing that the observed trends was explained with the assumption they made regarding the neutron capture nucleosynthesis (see Fig.\ref{fig2}).

\section{Cescutti et al. (2006)}

\begin{figure}[H]	
\widefigure
\includegraphics[width=10cm]{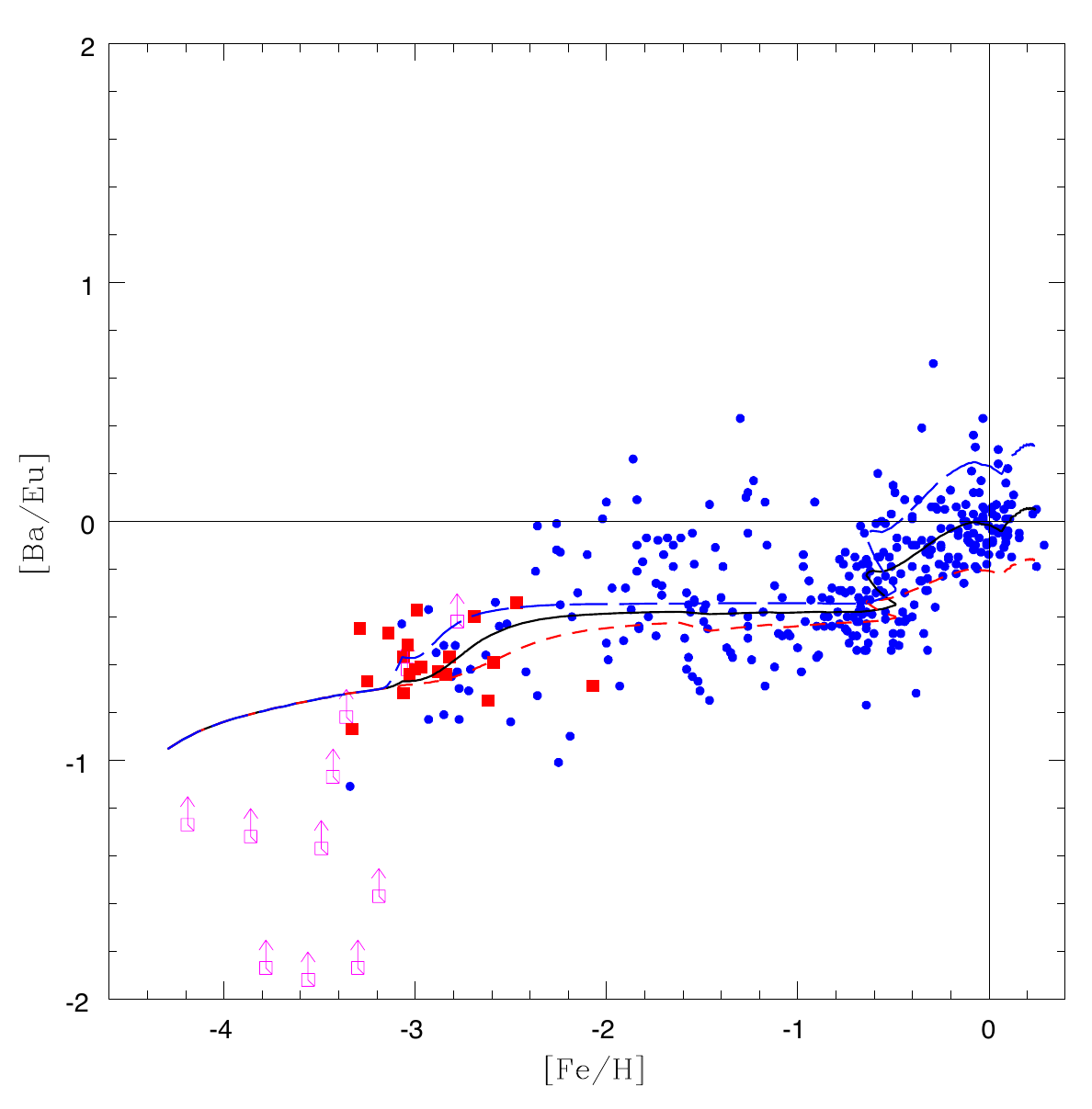}
\caption{ [Ba/Eu] vs [Fe/H]. Data (filled squares) and lower limits (open square) by \citet{Franc07}; in blue  observational data from \citet{Burris00,Fulbright00,Koch02,Honda04,Mashonkina00,Mashonkina01}. 
The solid line is the prediction of the best model 1,
short dashed line and the long dashed line are the predictions of max and min models, able to include most the observational data for [Ba/Fe] vs [Fe/H]. Image reproduced with permission
 from \citet{Cesc06}, copyright by the authors.}\label{fig3}
\end{figure}

The main goal of the work of \citet{Cesc06} to follow the evolution of
Ba and Eu abundances by means of a detailed chemical evolution model reproducing
already the abundance trends for other elements. The chemical evolution model  was similar to the "two-infall" model of \citet{Chiappini97}), {\bf where the halo plus thick disc form during a first fast gas infall episode, followed by a second longer infall event which forms the thin disc}. 
The main difference compared to the work of \citet{Travaglio99}, was a more detailed descriptions of the r-process production. 
It was concluded that for  both Ba and Eu, it was necessary to assume  an r-component originating in stars in the range 10–30 $M_{\odot}$. This outcome was obtained thanks to the careful comparison of the GCE models to the  results of the Large Programme led by R. Cayrel \citep{Cayrel04} and the high quality abundances obtained for neutron capture elements by \citet[][see Fig. \ref{fig3}]{Franc07}.
In particular, the observed [Ba/Eu] ratio in metal poor stars, at metallicities (epochs) when the enrichment by low mass stars has not yet occurred, appeared to be  slightly enhanced compared to the ratio of the r-process for [Ba/Eu]$\sim -$0.7, observed in r-process rich stars \citep{Sneden08}. In fact, the stars in the range$-$2$<$[Fe/H]$<-$1 have a mean [Ba/Eu]$\sim -$0.4, as also reproduced  by the model of \citep{Cesc06} assuming empirical yields and not so well by the model of \citep{Travaglio99}. This implies the necessity of a production of Ba by massive stars (or a source with a comparable timescale) slightly higher than the one expected for the pure r-process ratio. This source could be simply ascribed to a variation of the r-process for [Ba/Eu], but it can  be also connected to the production of  s-process Ba by low metallicity rotating massive stars \citep{Frisch15,LimongiChieffi18}.
This extra production of barium on short timescales has also a (moderate) impact on the required fraction of s-process barium coming from AGB stars, which is therefore slightly lower, decreasing from 80\% to 60\%. Later on, \citet{Cescutti07} applied the same assumptions for Eu and Ba adopted in the solar vicinity to model the Galactic gradients. The results were in excellent agreement with the observational data, with a steep gradient  for Eu and an almost flat gradient for Ba, due to the delayed enrichment by AGB stars. 

\section{D'Orazi et al. (2009) and Maiorca et al. (2011)}

\begin{figure}[H]	
\widefigure
\includegraphics[width=12cm]{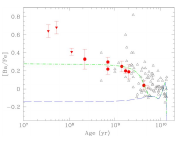}
\caption{Average [Ba/Fe] as a function of stellar age for the subsample of clusters whose analysis is based on dwarfs (filled circles and inverted triangles) compared with the abundance pattern of disk stars (open triangles) by \citet{Bensby05}. Filled triangles represent abundance measurements that probably need  NLTE (Non Local Thermodynamc Equilibrium) corrections {\bf \citep[but see][]{Baratella21}} The model results are shown for two set of yields: (a) standard yields \citep{Travaglio99,Busso01}, long-dashed curve; and (b) enhanced s-process yields, dot-dashed curve. Both models show a peak at old ages due to the r-process from massive stars. 
Image reproduced with permission
 from \citet{Dorazi09}, copyright by the authors.}\label{fig4}
\end{figure}

In \citet{Dorazi09}, it was reported the discovery of a trend of increasing Ba abundance with decreasing age for a large sample of Galactic open clusters (OC). The observed pattern of [Ba/Fe] versus age could be reproduced with a GCE model assuming a  Ba yield from the s-process in low-mass stars higher than the typical one suggested by parameterized models of neutron-capture nucleosynthesis
(see Fig. \ref{fig4}). They therefore showed that this is possible in a scenario where the efficiency of the extra-mixing processes, producing the neutron source $^{13}$C, anti correlates with the initial mass, with a larger efficiency for lower masses. This is similar to the known trend of extended mixing episodes acting in H-rich layers \citep{Charbonnel94} and might suggest a common physical mechanism \citep{Busso07}. 

Later on in \citet{Maiorca11}, the analysis of OCs  was extended to other elements besides Ba.
They derived abundances for  four other elements having important s-process contributions, i.e.  Y, Zr, La, and Ce. They used equivalent width measurements as well as the MOOG \footnote{http://www.as.utexas.edu/$\sim$chris/moog.html} code \citep{Sneden73} and their sample included 19 OCs of different ages, for which the spectra were obtained by the ESO Very Large Telescope using the UVES spectrometer. Thanks to these further abundance determinations,
they confirmed  for all the elements analyzed in their study, what was originally found only for barium. Their results  require
that very low mass AGB stars (M$\lessapprox$ 1.5 M$_{\odot}$) produce larger amounts of s-process elements (and hence activate the $^{13}$C-neutron source more efficiently) than previously expected. The role  of these stars in producing neutron-rich elements in the Galactic disk has been so far underestimated, and their evolution and neutron-capture nucleosynthesis should now be reconsidered.
{\bf The most recent observations have confirmed that the young clusters appear to be extremely enhanced in Ba young open clusters but the other s-process elements are not \citep[see e.g. recent work by][and references therein]{Baratella21}. This cannot be explained in terms of s-process nucleosynthesis and the current explanations for this Ba excess range from observational problems with Ba to the presence of a neutron capture process intermediate between s-process and r-process. }

\section{Rizzuti et al. (2019)}

More recently, \citet{Rizzuti19} presented a detailed study on the evolution of Sr and Ba in the Milky Way. These two elements have a predominant s-process origin and they adopted the yields from AGB stars in the range 1.5-3.0$M_{\odot}$ from \citet{Cristallo09,Cristallo11}. However, there is a minor s-process  contribution also from massive stars, and they tested the yields of \citet{LimongiChieffi18} and \citet{Frisch15}.  A small r-process component to the production of Sr and Ba was also taken into account by considering either SNe or merging neutron stars. 
\begin{figure}[H]
\widefigure
\includegraphics[width=13cm]{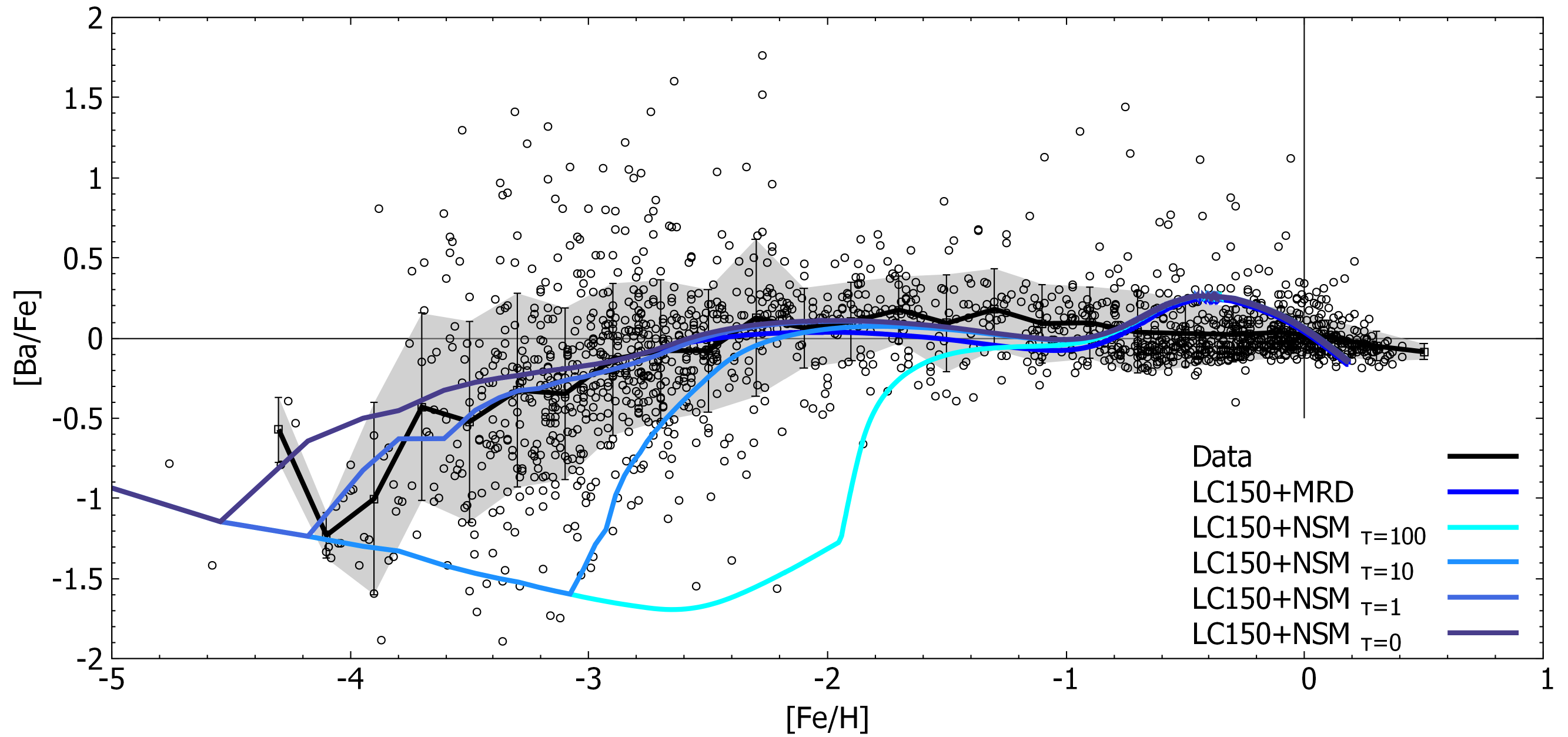}
\caption{Comparison between observed and predicted [Ba/Fe] ratios vs. [Fe/H]. The black dots, track, and shadowed area are the observations (sources listed in Table 1); dark blue line is model with magneto-rotationally driven SNe \citep[see][]{Thielemann11} as r-process sources; lighter blue line is the model with neutron star mergers as r-process source with variations in the time delay (from darker to lighter) with $\tau$ = 0, 1, 10, and 100 Myr. Image reproduced with permission
 from \citet{Rizzuti19}, copyright by the authors.
}
\label{fig5}
\end{figure}
As one can see from Fig. \ref{fig5}, the best model assumes the AGB yields from \citet{Cristallo09,Cristallo11}, the massive rotating star yields from \citet{LimongiChieffi18} with rotational speed of 150 Km/sec and the r-component of Ba from magneto-rotationally driven SNe (\citep[see][]{Thielemann11}).

\section{Grisoni et al. (2020)}

\begin{figure}[H]
\widefigure
\includegraphics[width=10cm]{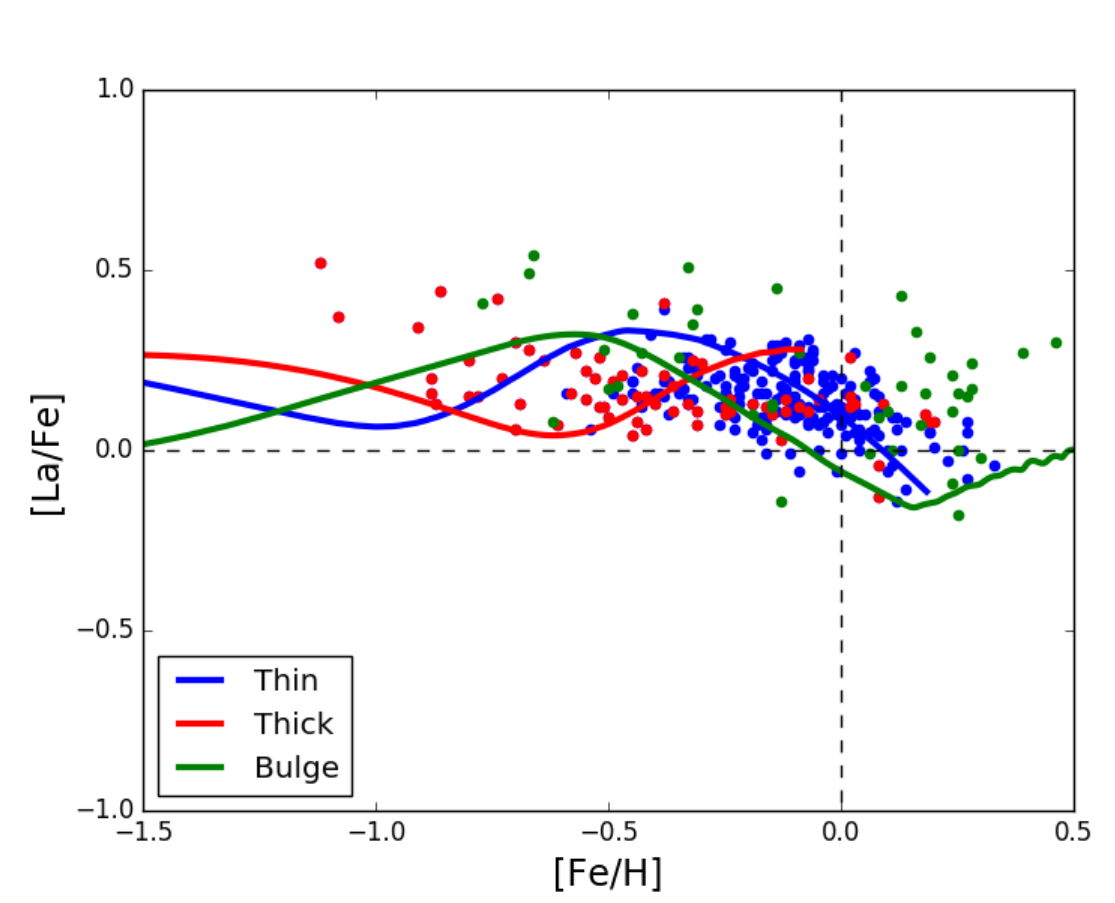}
\caption{Comparison between observed and predicted [La/Fe] ratios vs. [Fe/H]. The theoretical thick disk is shown by the red curve, the thin disk by the blue one, and the bulge by the green one. The data points for the three components are indicated by the same colors as the theoretical curves. Image reproduced with permission from \citet{Grisoni20}, copyright by the authors.}
\label{fig6}
\end{figure}

In \citet{Grisoni20}, the evolution of Zr, La, Ce, and Eu in the different components (thick, thin-disk and bulge)  of the Milky Way was followed. The models for the thick and thin disks  were from  \citet{Grisoni17} and the model for the bulge from \citet{Matteucci19}. Again, the yields adopted in \citet{Grisoni20} from s-process in AGB stars (1.3-3.0M$_{\odot}$) are from \citet{Cristallo09,Cristallo11}, together with the s-process contribution  by massive rotating stars from \citet{Frisch15}. As an example, we show in Fig. \ref{fig6} the predicted and observed abundances, relative to Fe, of La which is predominantly produced by the s-process, as opposed to Eu which is a pure r-process element.
As one can see, the agreement between theory and observations is good and this means that the stellar yields, which are the most important parameters in GCE models, are the appropriate ones. Moreover, the agreement ensures that also the assumed histories of star formation in the different Galactic components (thick, thin-disk and bulge) are the correct ones;
 in particular, the SFR in the Galactic bulge must have been very intense and have quickly consumed the gas, followed by the milder SFR in the thick and even more in the thin disk. These assumptions on the SFR in the three Galactic components ensure good agreement also for other chemical elements, such as $\alpha$-elements \citep{Grisoni17,Matteucci19}.

\section{Prantzos et al. (2020)}
\citet{Prantzos20} presented an interesting  method, based on GCE models, for assessing the s- and r- fractions of the Solar System abundances. They used accurate yields from low and intermediate mass stars as well as from rotating massive stars. 
\begin{figure}[H]
\widefigure
\includegraphics[width=10cm]{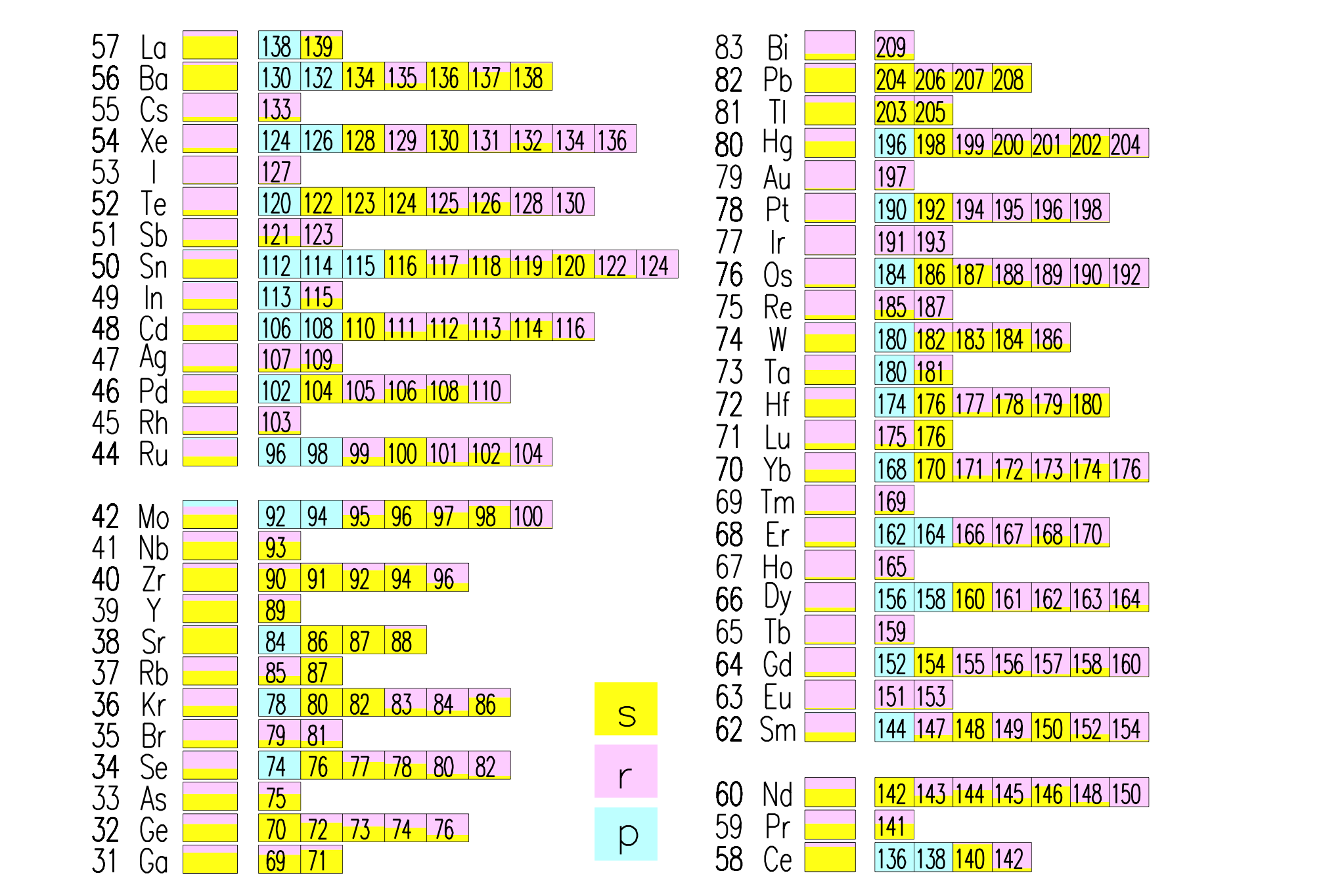}
\caption{Contributions of the s-, r- and p- processes to the solar chemical composition. The contribution of each process is proportional to the colored area of the corresponding box. Image reproduced with permission from \citet{Prantzos20}, copyright by the authors.}
\label{fig7}
\end{figure}
The method consists in running models with only one component (s- and r-) at the time. In Fig. \ref{fig7} we report the Table with the contributions from s-, r- and p-process to the heavy elements by \citep{Prantzos20}.

\section{Busso et al. (2021) and the s-Processing in AGB stars revisited adopting magneto hydrodynamics mixing}

In \citet{Busso21}, they assumed that magneto hydrodynamics (MHD) processes induce the penetration of protons below the convective boundary, when the Third Dredge Up occurs. There, the $^{13}C$ n-source can subsequently operate, merging its effects with those of the $^{22}$Ne($\alpha$, n)$^{22}$Ne reaction, activated at the temperature peaks
characterizing AGB stages.
\begin{figure}[H]	
\widefigure
\includegraphics[width=12cm]{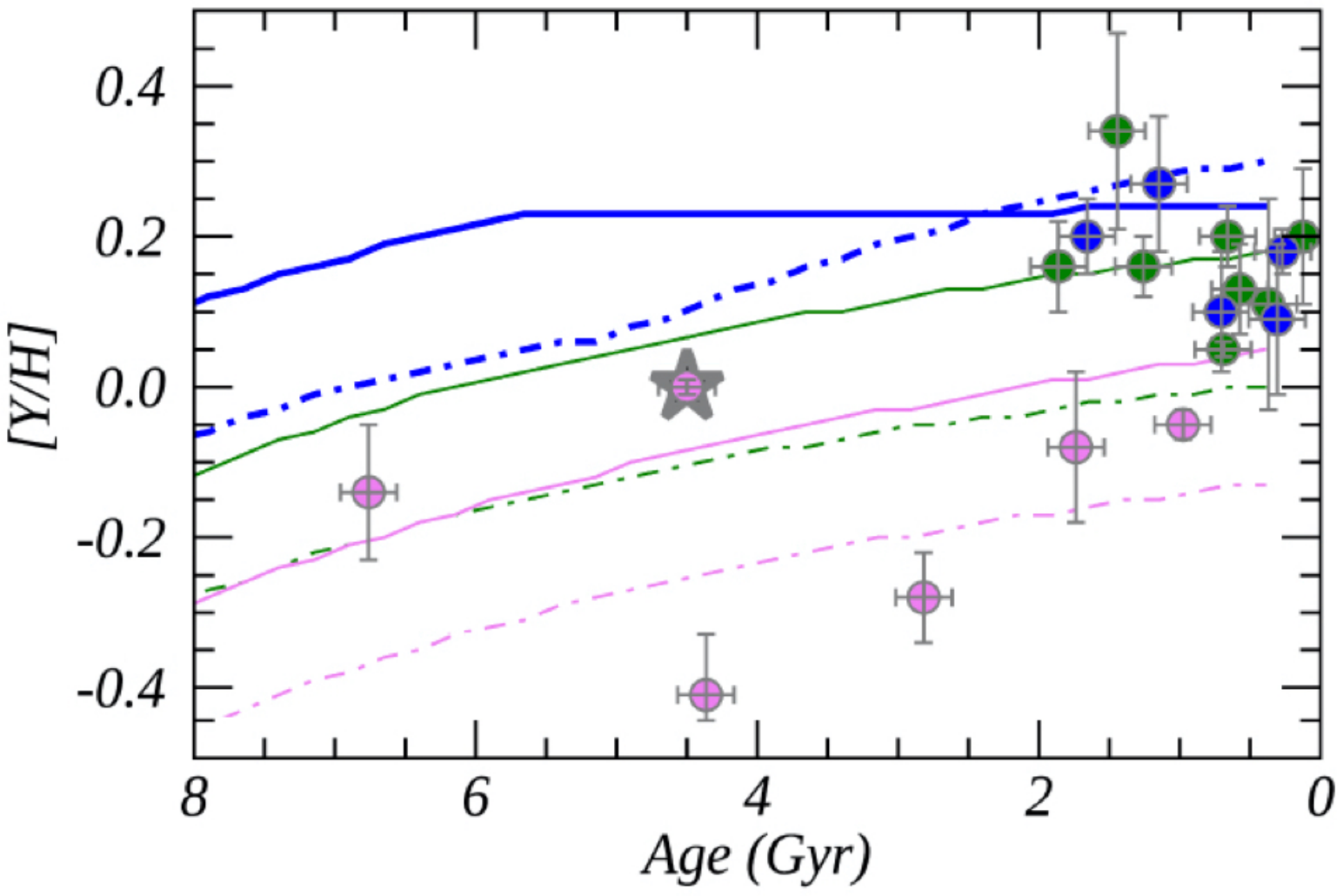}
\caption{[Y/H] vs. age: sample of Gaia-ESO idr5
clusters in three radial bins (Galactocentric radius (RGC) < 6.5 kpc in blue, 6.5 kpc < RGC <
9 kpc in green, and RGC > 9 kpc in pink) compared with results of the
GCE  model for the thin disk at three RGC = 6, 8, and 10 kpc with the 
yields computed with AGB models adopting mixing triggered by magnetic fields (continuous curves) and those with the FRUITY yields (dot-dashed
lines). The star marks the abundance ratio at the solar age and Galactocentric distance. Image reproduced with permission 
 from \citet{Magrini21}, copyright by the authors.
}\label{fig8}
\end{figure} 
 They also provided a grid of abundance
yields, as produced through their MHD  mixing scheme, uniformly sampled in mass and metallicity.  In that paper they showed that 
MHD-induced mixing is adequate to drive slow n-capture phenomena accounting for  observations of evolved stars and  isotopic ratios in presolar SiC grains {\bf \citet{Liu18}}. 
Moreover, in \citet{Magrini21}, these new yields were included in a GCE model and  the results compared with a sample of abundances and ages of OCs located at different Galactocentric distances.
It was shown that  the magnetic mixing causes a less efficient production of Y at high metallicity. Since a non-negligible fraction of stars with supersolar metallicity is produced in the inner disk, their Y abundances are affected by the reduced yields. 
Thanks to the results of the new AGB  yields based on MHD mixing, the GCE model of \citep{Magrini21} was able to reproduce the observed trends for  [Y/H] versus age at different Galactocentric distances, improving the outcome of previous AGB yields (see Fig. \ref{fig7}).

\section{Conclusions}
In this paper, we have  summarized the main results obtained in the last 40 years relative to the importance of the s-process production by AGB stars on the enrichment of heavy elements in our Milky Way. To do that, one needs to adopt a detailed GCE model taking into account chemical yields from stars and stellar lifetimes. By comparing model results and data on s-process elements measured in Galactic stars, it is concluded that  s-process elements such as Ba have a main s-process component produced in low mass stars and appearing on long timescales together with a faster component which has a r-process origin \citep{Cesc06}. Very likely this component originates on short time scales from massive stars or neutron star neutron star mergers \citep{Thielemann11,Perego21}. 
New detailed yields from AGB stars taking into account  MHD processes have also allowed to reproduce  recent data of Y  observed in open cluster stars.

In conclusion, by means of Galactic archaeology  together with more and more precise chemical yields (i.e. \citep{Busso01}, \citep{Busso95}, \citep{Busso21}), we are able to impose constraints on the origin of chemical elements as well as on the history of star formation in galaxies.

\acknowledgments{This work was partially supported by the European Union (ChETEC-INFRA, project no. 101008324).}

\reftitle{References}


\externalbibliography{yes}
\bibliography{biblio_busso.bib}

\end{document}